# Application of Business Intelligence In Banks (Pakistan)

Muhammad Nadeem and Syed Ata Hussain Jaffri
SZABIST
Karachi, Pakistan

***Abstract:*** *The financial services industry is rapidly changing. Factors such as globalization, deregulation, mergers and acquisitions, competition from non-financial institutions, and technological innovation, have forced companies to re-think their business.*
*Many large companies have been using Business Intelligence (BI) computer software for some years to help them gain competitive advantage. With the introduction of cheaper and more generalized products to the market place BI is now in the reach of smaller and medium sized companies. Business Intelligence is also known as knowledge management, management information systems (MIS), Executive information systems (EIS) and On-line analytical Processing (OLAP).*

## 1. DEFINING BUSINESS INTELLIGENCE

"Business intelligence is the process of gathering high-quality and meaningful information about the subject matter being researched that will help the individual(s) analyzing the information, draws conclusions or make assumptions." [Jonathan, DMR 2000]

Business intelligence refers to the use of technology to collect and effectively use information to improve business effectiveness. An ideal BI system gives an organization's employees, partners, and supplier's easy access to the information they need to effectively do their jobs, and the ability to analyze and easily share this information with others.

### 1.1. Traditional Reporting

Traditionally reporting in an organization often flows up the management hierarchy of the business e.g. Production operators will collect information about a production line, e.g. units produced, production time, down time and utilization %, this information will be passed to a shift supervisor who may well pass it in a summarized form to the production manager and then to a production director.

"The key to an information marketplace is an active information repository--or catalog--which contains or points to a variety of "information objects" both inside and external to the organization. Users can browse through the catalog, shopping for objects that interest them and publishing objects that they've created or modified for others to consume." [Eckerson, DMR 1998]

### 1.2. Analytic versus Business Intelligence

"Information workers at all levels of the organization need to be able to interact with the data: to drill down, drill up, slice and dice business information to quickly find the relevant facts on their own, without administrative intervention."[MS, MSW 2002]

Both business intelligence tools and analytic applications draw on information that has been sourced from multiple disparate systems across (and sometimes beyond) an enterprise and integrated into an information repository. Apart from this commonality, the contrast between them couldn't be greater.

Business intelligence tools have been likened to spreadsheets on steroids. They deliver powerful analysis and knowledge discovery capabilities into the hands of specialists who have gone through week-long training classes to become familiar in their use. But in practice, using a business intelligence tool is a painstaking and time-consuming process even for a power user. The user has to be proficient in the use of the tool, know how to structure ad hoc queries and SQL statements, and also understand how to perform intricate analyses. Consequently, with business intelligence tools, analysis is performed in a silo – separate from management functions rather than integrated with them. The resulting reports and forecasts are not always intuitively understandable and, in any case, represent after-the-fact knowledge.

In contrast, analytic applications are focused on immediacy of information, its broad deployment and its direct applicability across the entire enterprise value chain from front- and back-office operations to supply chain, CRM, Web channel, sales and marketing, and other critical line functions. Integration, analysis and delivery capabilities can all built into the application. Instead of waiting for reports to be sent to them by power users, analytics-enabled managers themselves use business problem-specific, Web- based dashboards and scorecards to evaluate key performance metrics on a continual basis. Rather than putting users through a voodoo-like process that is staggering in its complexity, analytic applications provide analytic workflows that guide managers quickly and consistently through their business decisions.

### 1.3. Strategic or Tactical

Business intelligence applications can be deployed either strategically i.e. across functional department or tactically i.e. within a functional department.

#### 1.3.1. Strategic

Strategic BI has the potential of big rewards. It can give senior managers a holistic view of the company and can identify trends and opportunities for growth. It can also be used for monitoring the company against its Key Performance Indicators (KPI's). Because it goes across departmental boundaries it encourages collaborative working in the organization.

### 1.3.2. Tactical

Can be applied to the 'pain' areas of your business where the extra knowledge and insight that BI can bring will bring quick and quantifiable results. It is usually a good place to start if you have had no previous experience in BI. An example of tactical BI deployment might be to look at production yield from a manufacturing process, we might want to record inputs, output, wastage, plant breakdown.

## 2. IDENTIFYING BUSINESS INTELLIGENCE OPPORTUNITIES

The first task in starting a BI initiative—and the first goal of the BI roadmap—is identifying what you want to achieve with business intelligence. In practical terms this means looking for opportunities in your organization where business intelligence can improve the quality of day-to-day decision making.

This process is divided into three primary steps:
1. Doing your homework: requires consideration of where business intelligence can be applied in an organization (for example, business units or functional areas), who is to benefit (for example, executives, analysts, and managers), and what types of information are needed (for example, dimensions and measures).
2. Sharing and collecting ideas: involves gathering people together to brainstorm and share their ideas and experiences about which business processes can benefit from business intelligence and what information can help them improve these processes.
3. Evaluating alternatives: uses standard criteria to assess the ideas collected during brainstorming efforts and identify those BI opportunities that offer the greatest benefits.

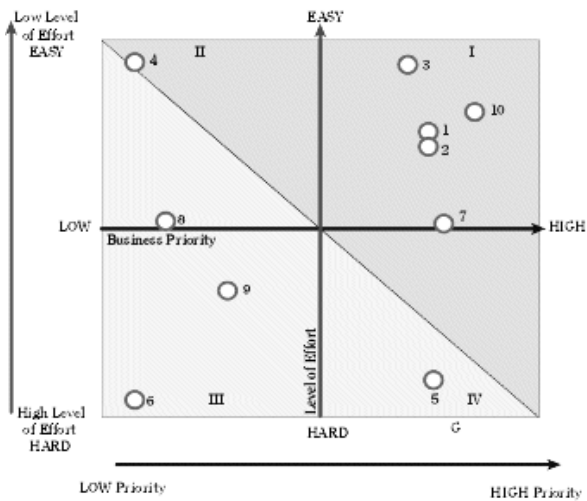

Figure 1. A sample BI opportunity scorecard [Source: Vitt Luckevich Misner, BI 2002]

### 2.1. Costs, Benefits & Returns

"Return on investment (ROI), the yardstick against which most corporate IT projects are measured, has not been consistently used as a justification for data warehousing for two reasons. First, in the rush to implement this highly popular decision support solution and important competitive weapon, early adopters have tended to evaluate data warehousing using less stringent criteria than for other technology outlays. Second, due to the relative immaturity of the technology, data warehousing projects are recognized as inherently risky and deserving of greater latitude in delivering ROI." [CRANFORD, DMR 1998]

BI opportunities are typically more difficult to evaluate than other IT projects using traditional return on investment, payback, and discounted cash flow techniques, especially for companies that have no experience with the technologies. OLTP systems are inextricably linked to the day-to-day processes of the business, where costs are generally well known and consequences of systems failures, for example, not processing an order or not invoicing a customer for goods shipped, are understood and easily quantified.

With business intelligence, however, the most important benefits, while intuitively obvious, are often not easily quantifiable in advance. They revolve around less measurable, more esoteric variables, such as the impact of having information sooner, the quality of decisions, new marketplace insights and tactics, and potential shifts in competitive strategy.

The list of intangible benefits, while difficult to quantify, is where the greatest and fastest paybacks occur.
- Improved operational and strategic decisions from better and more timely information
- Improved employee communications and job satisfaction resulting from a greater sense of empowerment
- Improved knowledge sharing

## 3. BUSINESS INTELLIGENCE INFRASTRUCTURE

Business organizations can gain a competitive advantage with a well-designed business intelligence (BI) infrastructure. Think of the BI infrastructure as a set of layers that begin with the operational systems information and meta data and end in delivery of business intelligence to various business user communities. These layers are illustrated in Figure below.

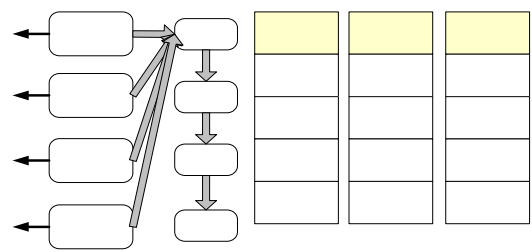

Figure 2 Business Intelligence Infrastructure's Layers [DEBROSSE, TD 2003]

## 3.1. Business Benefits

The payback achieved by building the business intelligence infrastructure is a function of how efficiently it operates, how well the infrastructure is supported and enhanced by the business organization as well as its capacity for producing business insight from raw operational data. The business intelligence infrastructure delivers key information to business users. For maximum impact, standards and procedures must be in place to provide key business information proactively. This business intelligence infrastructure enables the organization to unlock the information from the legacy systems, to integrate data across the enterprise and empower business users to become information self-sufficient.

Providing managers and knowledge workers with new tools allowing them to see data in new ways empowers them to make faster and better decisions. Rather than responding to continuous stream of report requests, the business intelligence platform provides business users self-service decision support via the Web or at the desktop.

## 3.2. Data Integration

"It's exactly this widespread source of data that has finance organizations struggling to meet the current challenges before them. As companies grow larger through mergers, acquisitions and global expansion, they collect and create more and more financial systems, a collection that becomes increasingly difficult to manage and integrate." [DEBROSSE, TD 2003]

Based on the overall requirements of business intelligence, the data integration layer is required to extract, cleanse and transform data into load files for the information warehouse. This layer begins with transaction-level operational data and meta data about these operational systems. Typically this data integration is done using a relational staging database and utilizing flat file extracts from source systems.

## 3.3. Information Warehouse

The information warehouse layer consists of relational and/or OLAP cube services that allow business users to gain insight into their areas of responsibility in the organization. Important in the warehouse design the definition of databases that provide information on confirmed dimensions or business variables that are true across the whole enterprise.

"What's needed to aggregate the data, then make it available to the appropriate decision makers is a data repository capable of pulling data from the disparate sources spread across the enterprise – an enterprise data warehouse (EDW)." [DEBROSSE, TD 2003]

In order to architect this information warehouse layer correctly, the business requirements and key business questions need to be defined. When this information is available, there will be additional insight into the business derived from the underlying data that cannot be fully anticipated before the data is actually available. Key areas to consider in defining requirements relate to the major functional areas of the organization.

## 3.4. BI Applications

The most visible layer of the business intelligence infrastructure is the applications layer which delivers the information to business users. Business intelligence requirements include scheduled report generation and distribution, query and analysis capabilities to pursue special investigations and graphical analysis permitting trend identification.

This layer should enable business users to interact with the information to gain new insight into the underlying business variables to support business decisions.

In order to achieve maximum velocity of business intelligence, continuous monitoring processes should be in place to trigger alerts to business decision-makers, accelerating action toward resolving problems or compensating for unforeseen business events. This proactive nature of business intelligence can provide tremendous business benefits.

## 3.5. Portals

Presenting business intelligence on the Web through a portal is gaining considerable momentum. Web-based portals are becoming commonplace as a single personalized point of access for key business information. All major BI vendors have developed components which snap into the popular portal infrastructure.

"Strictly focusing on the business intelligence (BI) aspects of corporate portals is dangerous because it misses the needs of end users." [KOUNADIS, DMR 2000]

## 3.6. How well is Your Business Intelligence Infrastructure Implemented and Supported?

To determine the completeness and adequacy of your BI infrastructure, answer the following questions. Any "no" answers indicate opportunity areas for improvement.

1. Do you have an effective data integration process in place to create required business intelligence on a daily basis?
2. Are continuous monitoring processes in place to allow alerts to be communicated immediately to those who need to take action?
3. Is your information delivery process automated?
4. Is your warehouse administration infrastructure completely automated?
5. Are alerting techniques used to communicate exceptions quickly so decisions are accelerated?
6. Are the key business questions being answered about your business areas of responsibility?

7. Is information available on standardized dimension such as customer, product and geography?
8. Do you have adequate competitive information to answer key business questions?
9. Have you delivered scorecards on key performance indicators to top decision-makers?
10. Do you leverage your enterprise portal infrastructure to deliver business intelligence?

## 4. BUSINESS INTELLIGENCE AND FINANCIAL INDUSTRY

The financial services industry is rapidly changing. Factors such as globalization, deregulation, mergers and acquisitions, competition from non-financial institutions, and technological innovation, have forced companies to re-think their business strategy.

"As competition intensifies in the retail financial services marketplace, accurate measures of customer value down to the account level are becoming increasingly pivotal to success at the retail end of the market. This applies both to established players and new entrants." [SIMON, NCR 2000]

Financial services companies now have to create new revenue streams, enter new markets, gain market share, and reduce operational costs.

In addition, customers' expectations are changing. They are becoming better informed and more demanding. Companies are therefore transforming their management strategy to become more customer-centric than product focused.

Though these challenges span the financial services industry, consumer banking, investment banking, and insurance each has its own unique demands that require different success strategies.

| Challenge | Process | Average | World Class |
|---|---|---|---|
| Provide timely, accurate information to decision makers. | Closing cycle | 5 - 8 days | < 2 days |
| Provide accurate forecasts to Investors | % of time on analysis/forecasting | 20% | 44% |
| Increase productivity and Bandwidth per function | FTEs per $B | 122 | < 95 |
| Control expenses | Cost of Finance | 1.15% of revenue | < .53% of revenue |

**Table 1: Financial Management Challenges**

*[Source: 2000 Hackett Group Benchmarking/Solutions Book of Numbers]*

Business intelligence solutions have played a significant part in the strategy of many of financial services' companies, to quickly adapt to market changes. With easy access to large amounts of complex data from disperse sources, companies are able to manage costs and performance, and acquire and increase the profitability of customers.

Business intelligence solutions can help financial services companies in many ways, including retail banking, insurance, and investment banking.

### 4.1. Retail Banking

Identify profitable customers and products by understanding customer buying patterns and characteristics.

Retain customers and increase their value by understanding individual customer behavior, identifying and responding to changing needs, and offering better products and services.

Optimize multi-channel interaction by understanding customer preferences, transaction costs, and channel performance.

Improve customer service by identifying and responding quickly to incoming call trends by customer, representative, geography, and dispute types.

Increase effectiveness of marketing campaigns by analyzing optimal response rates and channel strategies.

Reduce customer acquisition costs by identifying profitable customer characteristics and executing targeted marketing acquisition campaigns.

Reduce delivery costs by providing electronic bill reporting to enable customers to analyze their bills and make informed decisions.

Reduce risk and minimize losses by understanding risk exposure both throughout the organization and at an individual level.

"By establishing the value of each account to its business, the institution gets all the benefits of detailed data, plus the ability to aggre-gate this data in an infinite variety of ways. The value of this capability is reflected by the wide range of functions across the institution which queue up to make use of aggregated account data as soon as it becomes available." [SIMON, NCR 2000]

### 4.2. Investment Banking

"Increase customer satisfaction by exceeding demands of high net worth individuals through improved understanding of their needs, risk tolerance, and investment interests." [BO, BUSINESSOBJECT.COM 2003]

Improve customer loyalty and increase their value by providing access to account information so that customers can make faster and more effective company and employee investment strategies.

Respond quickly to meet regulatory requirements by reporting on the level of exposure of holdings and investments in volatile regions.

## 5. CIB – A BI IMPLEMENTATION

CIB-Credit information Bureau, a State Bank of Pakistan's department responsible for maintaining the information related to borrowing related to any person, company, and/or group of companies.

CIB maintains this information by frequently fetching borrowing's related information from various banks and institutes throughout Pakistan.

This is a requirement imposed by SBP on all financial institution that they need to get the credit worthiness report before granting a loan above a certain amount to a customer. The current practice of obtaining a credit worthiness report is that the Financial Institution submits a form in SBP. Here the report is prepared manually by the SBP staff and handed over to the requesting institution on the following day.

The reason behind to maintain such an information is to track the net amount hold by any borrower, to eliminate the manual work at SBP end, and to provide quick and easy service to the Financial Institution. This application will allow the user to access the Credit Information Bureau central repository in SBP.

### 5.1. Problem Statement

"To provide fast, accurate, and dynamic analysis on both individual borrower and Group basis."

Currently CIB gets borrowings information through fax and/or telephones and if any bank or institutes wants to inquire about the holding of a certain person it has to first contact CIB department and then after 15-20 days the information is provided to the bank.

The process is quite complex. CIB provide information about borrowings of a particular borrower, also if the borrower belongs to a particular group of companies then the groups net borrowing also needs to be identified.

### 5.2. Existing Repository

Currently the database is running on Oracle 9i. The database contains information related to borrowings and application security.

### 5.3. Entities Summary

Following is a summary of the operational data store. These tables will serve as source tables for our business intelligence application. In the table some description of each table is also specified.

Table 2: Operational Entity Summary

Out of 30 tables, in original schema, we will be using only 4 tables, as specified above, since the required analysis can be performed based on the information in these four tables.

### 5.4. The Solution

As with all business intelligence implementations, our solution is not just a software application. As discussed in previous sections a typical business intelligence solution consists of several layers. Starting from OLTP Data bases, to Data extraction, transformation, loading, generation of multidimensional data store, and finally a very user friendly User Interface providing Drill-down and Roll-up facilities.

As specified in the previous sections currently Oracle 9i has been used as data base of OLTP. In the following table the tools/technologies used in implementing each layer is specified.

| Layer | Tool/Technology |
|---|---|
| Operational Database (OLTP) | Oracle 9i |
| Data Extraction/Transformation/Loading | Microsoft Data Transformation Service |
| Data Staging Area | MS SQL Server 7.0 |
| User Interface | C#.Net, Pivot Table Service |

Table 3. Tool/Technologies Used

### 5.5. The Schema Design

We have used Star Schema in our data warehouse design. We have developed only one dimension for fulfilling all current analytical requirements.

We have only one fact table, for Borrowing detail information, and four dimension tables for Borrower, Institute, Director/Guarantor and Time related information.

### 5.6. Data Transformation

In order to extract data from Oracle 9i Server and then loads it into SQL server, we have used MS-SQL Data Transformation Services.

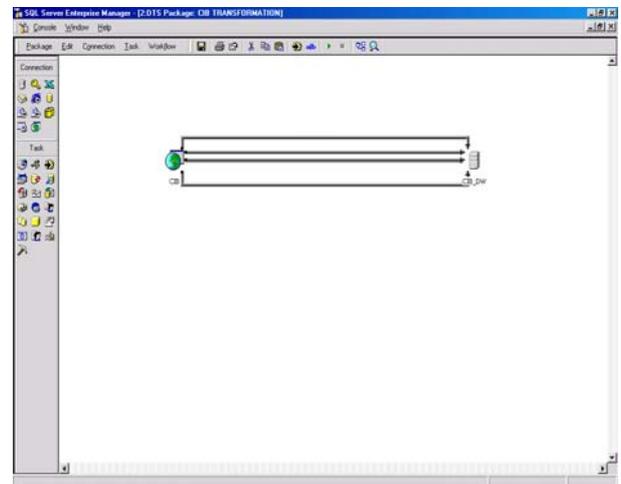

Figure: Data Transformation, Extraction and Loading

### 5.7. The Interface

The User interface is developed using Visual Studio.Net (C#.Net). By default user interface for any analytical/Business Intelligence Application needs to be dynamic and very user friendly.

OLBA-Online Business Analysis – is a very user friendly and easy to use application. It is very secure, dynamic and analytically enriched tool. OLBA provides graphical as well as analytical analysis to user on subject areas according to its rights/permission. OLBA allows user to perform analysis on various basis. For example, through OLBA user can analyze net borrowings based on Borrower, Institute, Time, on director/guarantor. Also

OLBA allows user to view aggregation of facts on various levels as designed in the Data warehouse Cube.

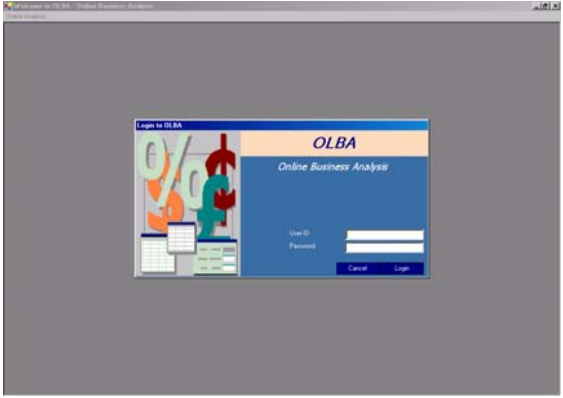

Figure: OLBA – User Interface

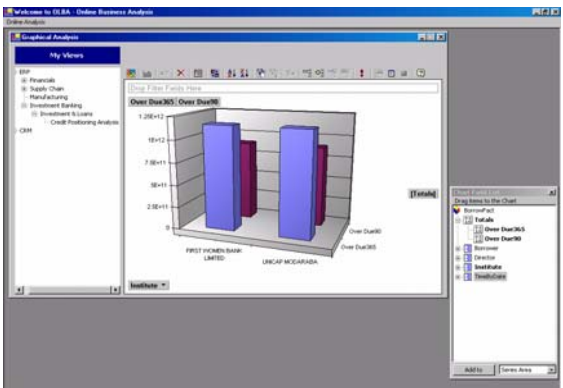

Figure3: OLBA – Graphical Analysis

Figure: OLBA – Numerical Analysis

## ACKNOWLEDGEMENTS

I wish to express my appreciation to Mr. Muhammad Nadeem, MSCS / MCS Coordinator of Shaheed Zulifqar Ali Bhutto Institute of Science and Technology. Who has always been a source of encouragement and knowledge for me, guided me in every step, and shared his knowledge. He has been most generous and understanding with his time to read this thesis carefully and make insightful comments and suggestions.

## REFRENCES

[Jonathan, DMR 2000] Jonathan Wu, February 2000, Business Intelligence: What is Business Intelligence?, DM Review.

[MS, MSW 2002] Microsoft, October 2002, Delivering Business Intelligence to the Enterprise using Microsoft Office-based Solutions, Microsoft (http://www.microsoft.com/office/business/intelligence).

[CRANFORD, DMR 1998] Stephen Cranford, January 1998, Knowledge Through Data Warehousing: Measuring, Managing and Retaining ROI, DM Review.

[DEBROSSE, TD 2003] Mike Debroose, 2003, Getting to the Bottom Line, NCR Corporation.

[KOUNADIS, DMR 2000] Tim Kounadis, February 2000, Business Intelligence for Intelligent Business, DM Review.

[SIMON, NCR 2000] Simon Doherty, 2000, CUSTOMER VALUE: GAINING THE COMPETITIVE EDGE, NCR.

[BO, BO 2003] www.BusinessObjects.com